\documentstyle[aps,multicol]{revtex}

\begin{document}

\title{Quantum Triangulation and Violation of Conservation of Trouble}
\author{Terry Rudolph}
\address{Department of Physics, University of Toronto,\\ Toronto M5S 1A7,\\ Canada}  

\date{\today}

\maketitle

\begin{abstract}
A scheme is described that allows Alice to communicate to Bob where on earth she is, even though she doesn't know herself. The situation described  shows how the generalization of a recent result of Gisin and Popescu [quant-ph/9901072] could be useful. 

\end{abstract}


\def\>{\rangle}
\def\<{\langle}
\def\be{\begin{equation}}
\def\ee{\end{equation}}
\def\bea{\begin{eqnarray}}
\def\eea{\end{eqnarray}}
\def\ba{\begin{array}}
\def\ea{\end{array}}
\def\nn{\nonumber}

\begin{multicols}{2}

Consider the following (rather contrived) scenario: Bob's friend Alice has been abducted. She regains consciousness to find herself locked in a windowless room, presumably somewhere on Earth. Checking her pockets she finds that she still has her mobile phone, and so she calls Bob. After being reassured that she's unhurt, Bob naturally asks ``Where are you?''. Of course Alice has no idea, and they discover that the phone company is singularly unhelpful and refuses to co-operate in Bob's attempt to locate Alice. (This part of course is the most realistic part of the story). Fortunately Alice realizes that she still has in her pocket the box of $N$ spin-$\frac{1}{2}$ particles that are entangled (in singlet states) with particles in Bob's possession. The question we are interested in is how may these particles be used to locate Alice?\footnote{We could also assume that Alice has a  bathtub (or toilet) in her room, and she can therefore inform Bob as to which hemisphere she is in.}

The simplest solution is the following: Alice alignes her mini Stern-Gerlach device (which will soon be standard on all Swiss-Army knives) in {\it her} vertical direction (${\bf A}_z$). She then proceeds to measure the spin of each of her particles in this direction, announcing to Bob over the phone what the result ($\uparrow,\downarrow$) of each measurement is. Bob is now left with $N$ particles, approximately one half of which are in the state $|\uparrow_{{\bf A}_z}\>$ and the remainder in $|\downarrow_{{\bf A}_z}\>$. If he can determine the direction ${\bf A}_z$, then he can determine where on Earth Alice is. Of course since they only share a finite number of particles he will only be able to do this to a certain accuracy.

The question to be addressed, therefore, is what measurement strategy should Bob follow in order to obtain the best estimate as to the direction ${\bf A}_z$? One strategy is to consider the $N$ particles as two different sets of particles with paralell spins, and to apply the procedure of [1] to each ensemble seperately. However Gisin and Popescu recently showed in [2], that a pair of anti-parallel spins is in fact preferable to a pair of parallel ones, given the task of determining as accurately as possible the direction the spin is pointing! If the result generalizes (and the purpose of this note is to motivate addressing this question), then in fact Bob would do better to apply a joint measurement to all $N$ particles, which is not so surprisng. But what is really amazing is that he would do {\it better} than if Alice had been allowed to send him a carefully prepared box of spins all pointing in the same direction. The randomness of the outcomes of her measurements is actually helping the task at hand! This appears to be a clear violation of the Law of Conservation of Trouble. (While no precise formulation of the Law exists in non-relativistic Quantum Mechanics, the Quantum Field Theory version can be found in [3], and of couse we are all familiar with the Law in the macroscopic limit, whence it is more often called Murphy's Law.)

In conclusion therefore; a fairly pithy motivation for determining the optimal measuring scheme for $N$ spins, about half of which are anti-aligned to the others, has been presented.

\end{multicols}

\end{document}